\documentclass[conference]{IEEEtran}
\IEEEoverridecommandlockouts
\usepackage{cite}
\usepackage{amsmath,amssymb,amsfonts}
\usepackage{algorithmic}
\usepackage{graphicx}
\usepackage{textcomp}
\usepackage{xcolor}
\usepackage{times}
\usepackage{helvet}
\usepackage{courier}
\usepackage{graphicx}
\usepackage[ruled,vlined]{algorithm2e}
\usepackage{color}

\definecolor{pink}{rgb}{0.858, 0.188, 0.478}

\def\BibTeX{{\rm B\kern-.05em{\sc i\kern-.025em b}\kern-.08em
    T\kern-.1667em\lower.7ex\hbox{E}\kern-.125emX}}
\begin{document}

\title{ULTRA: A Data-driven Approach for Recommending Team Formation in
Response to Proposal Calls}

\author{\IEEEauthorblockN{Biplav Srivastava}
\IEEEauthorblockA{\textit{Artificial Intelligence Institute} \\
\textit{University of South Carolina}\\
Columbia, USA \\
biplav.s@sc.edu}
\and
\IEEEauthorblockN{Tarmo Koppel}
\IEEEauthorblockA{\textit{Organization and Management Group} \\
\textit{Tallinn University of Technology}\\
Tallinn, Estonia \\
tarmo.koppel@taltech.ee}
\and
\IEEEauthorblockN{Sai Teja Paladi}
\IEEEauthorblockA{\textit{Artificial Intelligence Institute} \\
\textit{University of South Carolina}\\
Columbia, USA \\
spaladi@email.sc.edu}
\and
\IEEEauthorblockN{Siva Likitha Valluru}
\IEEEauthorblockA{\textit{Artificial Intelligence Institute} \\
\textit{University of South Carolina}\\
Columbia, USA \\
svalluru@email.sc.edu}
\and
\IEEEauthorblockN{Rohit Sharma}
\IEEEauthorblockA{\textit{Artificial Intelligence Institute} \\
\textit{University of South Carolina}\\
Columbia, USA \\
rohits@email.sc.edu}
\and
\IEEEauthorblockN{Owen Bond}
\IEEEauthorblockA{\textit{Artificial Intelligence Institute} \\
\textit{University of South Carolina}\\
Columbia, USA \\
obond@email.sc.edu}
}

\maketitle

\begin{abstract}
We introduce an emerging AI-based approach and prototype system for assisting team formation when researchers respond to calls for proposals from funding agencies. This is an instance of the general 
problem of building teams when demand opportunities come periodically and potential members may vary over time. 
The novelties of our approach are that we: 
(a) extract technical skills needed about researchers and calls from multiple data sources and normalize them using Natural Language Processing (NLP) techniques, (b) build a prototype solution  based on matching and teaming based on constraints, (c) describe initial feedback  about system from researchers at a University to deploy, and  (d) create and publish a dataset that others can use.
\end{abstract}



\section{Introduction}


Building teams in response to an opportunity is a common business activity. Examples are: responding to calls for proposals in product and services supply chains, expert teams for a medical procedure at a hospital, players for a match for team-based sports and crew for an airline flight.
In this paper, we will focus on teaming for researchers applying to funding agencies in response to their call for proposals, denoted {\em TeamingForFunding}.

A large proportion of funding for research in public universities comes from funding agencies. Hence, it is very important for researchers to be able to identify funding opportunities and make successful proposals.
Moreover, many of the funding opportunities are multi-disciplinary requiring teams to be quickly assembled from a wide variety of backgrounds who can work together. 
The advantage of this setting is that all data is readily available in public - proposal calls and their successful decisions (awards)  from agencies like National Science Foundation (NSF) and National Institutes of Health (NIH), and profiles of researchers  from data sources like Google Scholar. The users interested in the problem are researchers who will like to collaborate as well as administrators at researchers' organizations (e.g., Universities) who want to promote more collaborations, proposals and diversity at their institutions.
The objectives of teaming  can be short-term ($S$), long-term ($L$) or a combination of the two. 

 \begin{figure}[h]
   \includegraphics[width=.9\linewidth]{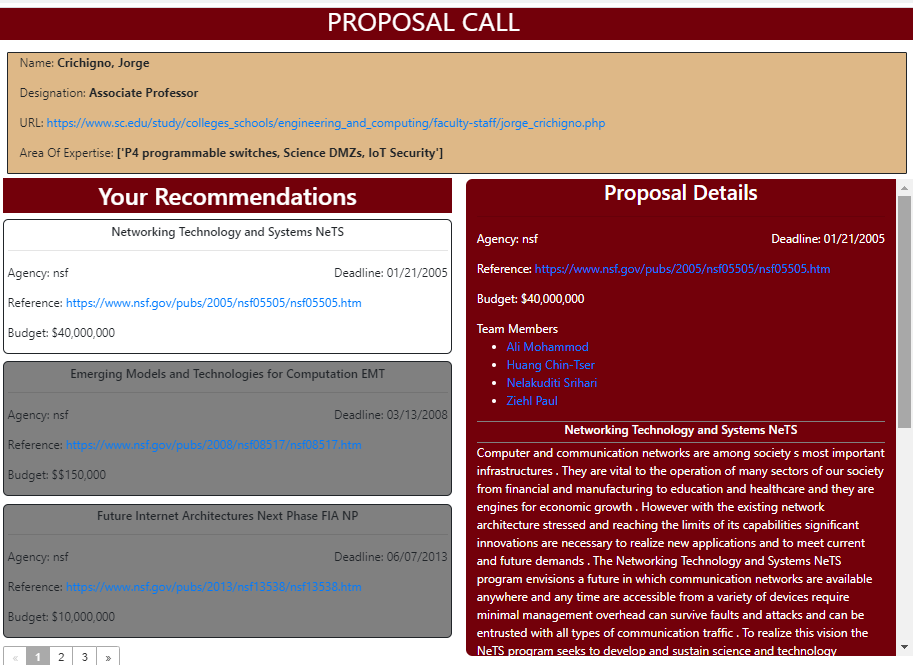}

 \caption{\footnotesize Team participant view of ULTRA showing team recommendations for select proposals (user view).}
 \label{fig:user-login}
 \end{figure}

\begin{figure*}[h]
    \includegraphics[width=\linewidth, height = 6cm]
    {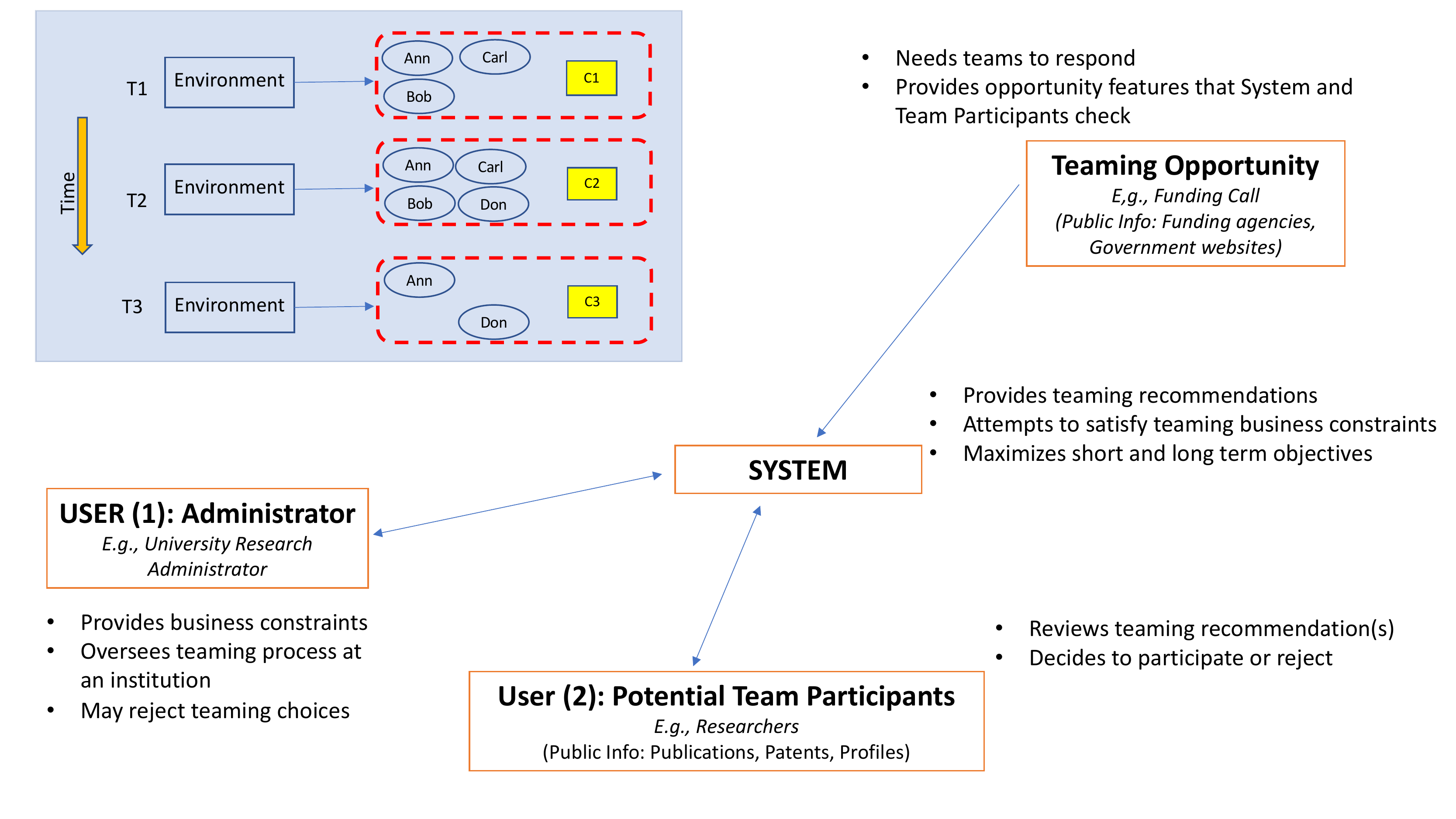}

\caption{\footnotesize Team-building setup with users who can be administrators or potential team participants.  A trusted team is one where all stakeholders are convinced that the right members from those available have been selected on the needs of the {\em call} for unknown success in future.}
\label{fig:setup}
\end{figure*}


In this paper, we present a novel approach and prototype  system called ULTRA\footnote{ULTRA stands for  University Lead Team builder from Rfps and Analysis.}. The system extracts technical skills from proposal calls found at multiple open data sources and those present in profiles of researchers online, and normalizes them using Natural Language Processing (NLP)  techniques. It then performs matches between calls and potential researchers, and from these matches, creates teams respecting teaming constraints. The recommendations can be  informed based on features from recent awards if the calls are repeating, and skill normalization extended using technical classification trees/ ontologies. 
Initial feedback of ULTRA from researchers at a University has been promising and in future, we plan a large scale evaluation of the system for a large administrative unit (College).
We have created a dataset of calls, profiles and recommendations that others can use as baseline and extend.

In the rest of the paper, we first give a demonstration of the current prototype and then describe background of the problem and related work. Then we formally introduce the problem and describe our solution and implementation. We then present initial evaluation and conclude with a discussion of future work.

\section{A Demonstration}


In Figure~\ref{fig:user-login}, we show how the system works for an individual user who can become a team participant. 
When the user has logged in, he can see his details on top panel like {\em Name}, {\em Area of Expertise}, {\em  Designation}. The Area of Expertise - {\em P4 programmable switches}, {\em Science DMZs}, {\em IoT Security} -  have been generated from the extracted research interest from the researcher's personal website and the Google Scholar account.

Below researcher's details, the system shows a list of recommendations on the left, 3 at a time (a configurable parameter, and details of a selected recommendation on the right. 
In each proposal, the system  shows the proposal call and team members recommended to be in the  team, an estimated budget,  and details from the proposal call like deadline and description. 
In Figure~\ref{fig:user-login},
the first recommendation is for {\em Network Technology and System NeTS Proposal}.
The system displays some important
information like the Agency name, link (URL) to the call and the deadline. It also shows
the supporting team members ({\em Ali Mohammod}, {\em  Huang Chin}, {\em Nelakuditi Srihari}, {\em Ziehl Paul}),
whose expertise complement the user as lead and a proposed budget. 

A few technical challenges are apparent in this example. The public data about researchers background may be obsolete or not  reflect their future research interests, the technical terms used in the proposal call and researchers background may not match, the budget size allowed in the call has to be respected while recommending team participants,  the recommended participants 
may already know each other but may not be free to work on the proposal. The recommendation from ULTRA can be seen as a  data-based trigger to kickstart collaboration transparently at an institution.


\section{Teaming for Funding Domain and Problem}

In this section, we provide preliminaries about the domain and introduce the problem.
Then in the next section, we describe the related work.



Funding agencies issue {\em Request For Proposals} (RFPs) on themes where they are looking for ideas which they can fund. We will also refer to RFPs with the term {\em calls}.  Researchers respond to RFP calls with proposals where they explain their ideas, list the activities that will be conducted and budget to complete the work.  They look to team with other colleagues to respond to such calls.

The team building setting we consider is shown in Figure~\ref{fig:setup}. Here, on the left, at different times, the environment offers candidate participants (indicated by blue ovals) the chance to match specific opportunity (indicated by square). The candidate set may change over time with significant overlap across time periods and also their skills or interests. On the right, the interactions within the proposed system at each time period are shown. The users of the system can be administrators representing institution(s) ($U_a$) as well as candidate team participants ($U_p$).  
They interact with the system during an opportunity ($T_i$) at time $i$. The system creates teaming choices which the users accept or reject.

At a given time, the inputs for a {\em TeamingForFunding} problem are: (a) a call for request for proposal (called call or RFP for short), (b) the profiles of researchers at an institution who are eligible to participate in teams, and (c) business objectives of the administrator and team participants at the institution. The solution to  the teaming  problem is a list of candidate teams where each team has two or more members with one designated leader. 
Optionally, each team will have an estimation of team’s budget and chance of the team's  success chances.

The objectives of teaming  can be short-term ($S$), long-term ($L$) or a combination of the two. 
For example, in {\em Teaming For Funding}, the short-term aims can be:  meet skill needs of funding opportunity ($S^f_a$), increase chance of success ($S^f_b$), satisfy business constraints of researchers' institutes ($S^f_c$)\footnote{Examples of business constraints: not over-burden the experienced researchers whose historical performance data is available, give opportunities to new researchers with little historical data, avail special provisions in funding opportunity.}. In the long-term, the objectives can be: maximize awards size over a time period ($L^f_a$), have a robust (diversified) pipeline of experienced talent ($L^f_b$), and satisfy diversity goals of researchers' institutions ($L^f_c$).

Further, teaming solutions consist of two phases: 
(a) \emph{matching}, to determine which researchers may be of interest to the calls based on skills needed, and (b) \emph{grouping}, to determine which subset of researchers should be recommended to be in a team. We address both these phases.

\section{Related Work}

We now discuss previous work related to the {\em TeamingForFunding} problem which are along teaming and matching.

\begin{figure*}[h]
    \includegraphics[width=\linewidth, height = 6cm]
    {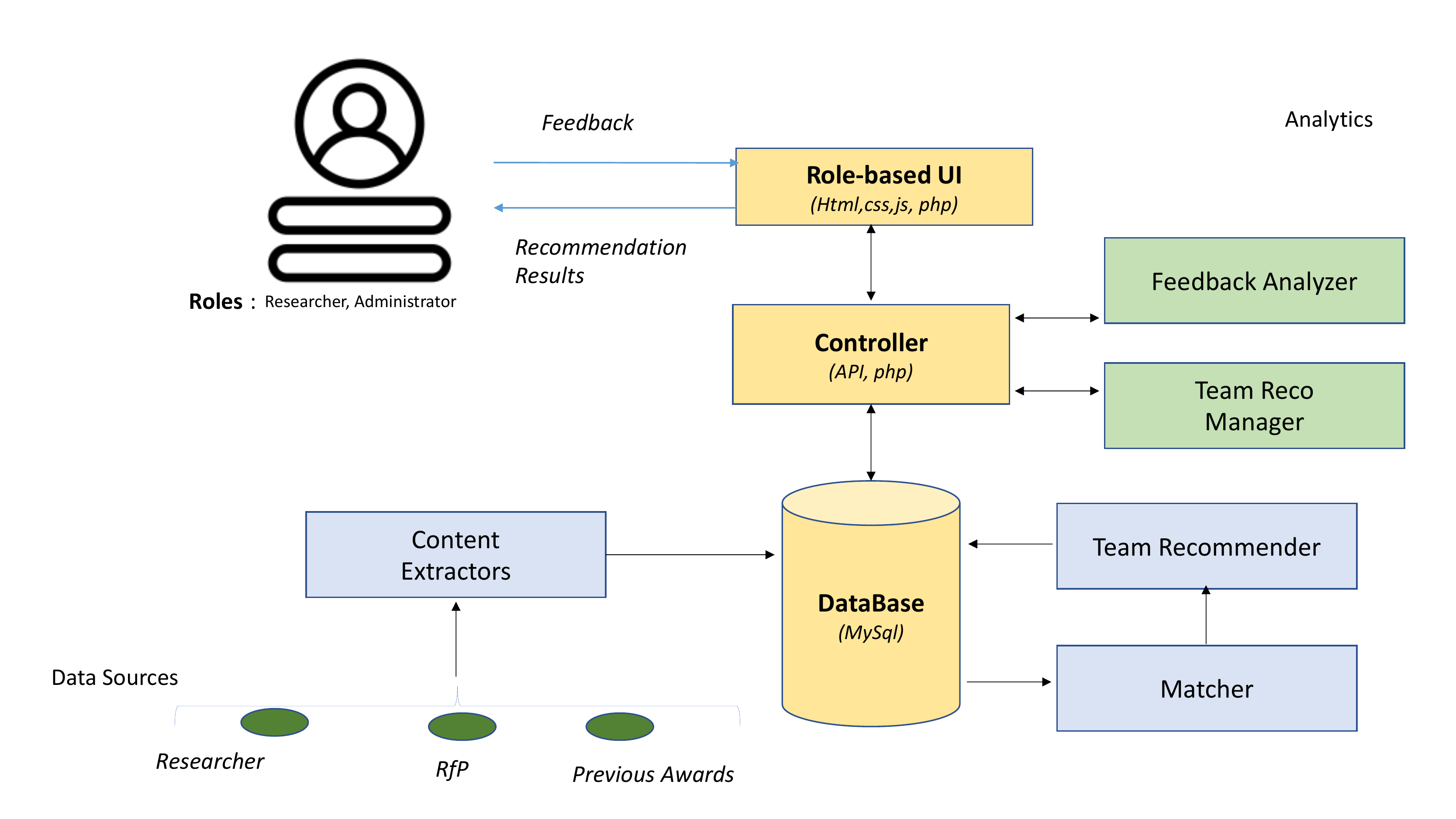}

\caption{\footnotesize System Architecture of ULTRA.}
\label{fig:sys-arch}
\end{figure*}

\noindent {\bf Team Formation:}
There is a rich history of AI in team formation. The most well studied type  of  teams are those formed in the {\em Hedonic Games} framework \cite{hedonic-games} where a team member only cares about other team members in their group. In our setting, a type of non-participants,  administrators at institutions,  are considered. However, our setting does not have another type of non-participants - user of a team's output like a patient in medical teaming. In \cite{teaming-optimiz}, the authors consider the problem of creating  general teams of {\em equal sizes}. 

Team formation has also been considered in sports team selections. In~\cite{ahmed2013multi}, the authors used a systematic, multi-objective genetic algorithm and multiple criteria decision making aids to optimize team selection for cricket tournaments. Some factors that the system takes into consideration are batting and bowling performance of each candidate player, brand value of players, success rate of the captain, wicket-keeper's past performances, and overall team budget. Based on the team selector's requirements, each of the above factors are given a certain ranking and the final formation set is picked according to it. If all the factors are deemed equally important, a domination approached can be applied instead. \cite{forbes2006entrepreneurial} theoretically explore team formation in entrepreneurial settings, i.e., studying how the addition of a new member to a team may impact overall performance. From prior literature, the authors summarize two views for team member additions. The first view identifies that new members are typically added to a team to fill certain resource and economic needs (e.g., hiring solely based on required skill sets). The second view states that adding a new member might impact or be driven by existing team members' social psychological needs. This means that new members may be added by relational trust, homophily (e.g., with respect to personal characteristics such as gender, race, and ethnicity), and similarities personality trait characteristics. Though these two views seem contrasting, the authors argue that they are not mutually exclusive, and further advise that instrumental needs coupled with personal relationships may increase a team's overall success within a venture.
\cite{gatson2005agent} explored dynamic team formation in the context of multi-agent systems (e.g., supply chains, sensor networks). They introduced the idea of agent-organized networks (AONs) and identified two robust strategies to improve organizational performance in dynamic supply and demand environments. In the first strategy, agents adapt their network connectivity based on \emph{preferential attachment}, a concept derived from graph theory which states that the more connected a node is within a network, the more likely it is to receive newer connections. The second AON strategy is based more on performance and referrals. Agents first derive an estimate of their individual performance, and then form connections with their highest-performing neighbors. Agents whose performance estimates are lower than the average of that of their neighbors' are removed. This strategy especially allows for real-time network learning and leads to more efficient network structures.

In recent literature,~\cite{juarez2021comprehensive} developed a comprehensive generalization of team formation problems and identified various applications for which team formation algorithms are formulated. They may also be used in domains such as software development, relief teams in the events of natural disasters, and cabinet formation in political settings, and search-and-rescue operations in hostile conditions. Due to the various factors and constraints (e.g., budget, time), team formation problems can quickly become NP-hard. The authors of this paper discuss popular techniques used in earlier research. Genetic algorithms have been used to \emph{evolve} better teams, i.e., gathering and optimizing collective abilities of team members rather than individual ones. Other methods include heuristics-based approximations, which reduce computational complexity by first gathering top-\emph{k} highest suitable candidates and pruning the rest. Note that \emph{k} is a small value here but still high enough for an optimized solution to exist. Clustering is also another popular technique, where teams are formed by grouping candidates with similar cost, team suitability scores, and total interaction scores~\cite{hlaoittinun2007team}.

In some unique scenarios, such as negotiation teams in hostage crises, teams may be required to form in the event of loss or unavailability of some members. Cases like this require robust and reliable formulation algorithms. \cite{okimoto2015form} proposed a community-based team formation problem, which calculates robustness scores (a non-negative integer) for potential teams and also takes member skill redundancy into account. A team that can function and meets skill requirements even with the removal of \emph{k} members is defined as \emph{k}-robust. The authors also set a cost restriction \textit{c}, thereby making it a bi-objective constraint optimization problem.

\cite{rakesh2016probabilistic} explored team formation in crowdfunding and used a probabilistic generative model to build \textit{CrowdRec}, which recommends Kickstarter projects to
a group of investors by accounting for each project's current status, future plans, preferences of individual members, and collective preferences of the group.

We recognize teaming as an ongoing area of research and propose to advance the state-of-the-art in fair teaming with a wider set of stakeholders inspired by practical applications.

\noindent {\bf Matching Concepts:}
Matching of concepts is a well studied problem in Computer Science and AI \cite{text-matching-survey,entity-matching}. There are two variants - matching the concepts as expressed representations of strings \cite{text-matching-survey} or as real world entities \cite{entity-matching}. We presently consider concepts as strings during matching.

There are many techniques to match strings starting with exact matches. Another method is
Fuzzy String Matching, also known as approximate string matching, to
match a pattern approximately within a distance threshold rather than exactly.
This allows us to find matches even when users misspell words or provide partial words for the (skill) search.
Another type of technique is embedding based matching. Here, a (deep) learning architecture learns a word or document level representation (embedding) from a large document corpus and numeric operations, like cosine similarity, is used to find similar terms \cite{embedding-survey}. Another type of technique, popular in recommendation literature, relies on multi-arm bandits techniques when matching happens repeatedly over time and with different users \cite{multi-arm-intro}.

\noindent {\bf Commercial Software for Advertising Researching Funding:}

For research funding, there are a few commercial offerings.
One system is Pivot \footnote{https://exlibrisgroup.com/products/pivot-funding-opportunities-and-profiles/} which sends keyword based alerts to faculties whose
interests match the areas listed in a RFP. Another system is Scry\footnote{https://www.amaforge.com/} which matches proposals to
faculty but does not identify teaming opportunities. 
In focus sessions with researchers, we learnt that  researchers wanted to have decision support tools that can help them with knowing more about calls, fellow researchers and about teaming opportunities.








\section{Solution}

We show the main components of our solution
in Figure~\ref{fig:sys-arch}. They consist of a Content Extractor (CE) to consume public data about researchers, calls (RFP) and previous awards. We have a Matching system to match calls to researchers (M)  and a Team Recommendation (TR) system to help for system output. The system also has components to analyze users feedback of output (FA) and a team recommendation manager (TRM) to send notifications to candidate team participants for their confirmations and helping the team prepare the final submission. 
We now explain the working of the main components.

\subsection{Content Extractor (CE)}

The CE module extracts content through webpage or Application Programming Interface (API) requests. It utilizes  {\em BeautifulSoup}\footnote{https://pypi.org/project/beautifulsoup4/} 
library to parse through exhaustive  content that may sometimes span hundreds of pages. It also uses the
{\em Spacy}\footnote{https://spacy.io/} library  to extract key details such as exact budget of RFP.

\begin{table}
\centering
    \begin{tabular}{|p{2.0cm}|p{2.0cm}|p{2.0cm}|}
    \hline
          Type & Number & \% Extracted  \\ \hline \hline
     RFP & 1,797 & 100   \\ \hline
     Title & 1,782 & 99.1   \\ \hline 
     Deadline & 1,626 & 90.4   \\ \hline
     Budget & 1,729 & 96.2   \\ \hline
     Synopsis/ Keywords & 1,797 & 100  \\ \hline \hline 
     Users & 240 & 100 \\ \hline
     Users/Research & 205 & 85.4 \\ \hline 
    \end{tabular}
    \caption{Performance of content extraction for calls and users. Users/Research refers to number of users we were able to confirm as researchers with a match with Google Scholar.}
    \label{tab:extr-results}
\end{table}

RFPs are extracted from the NSF website's search page\footnote{https://www.nsf.gov/publications/index.jsp}.
We use data about both historical calls whose deadline has expired and also open calls whose deadline is in future.  
From the search page, we acquire links to the individual RFP pages, then extract the HTML and parse it for important information such as the synopsis, title, and application deadlines. 
Table~\ref{tab:extr-results} gives the statistics of extraction for each information type. The extractors are generally accurate but they face challenges with calls when they have multiple sub-tracks and corresponding deadlines.

We also looked at XML archives hosted by NSF, and the search page of Grants.gov\footnote{At: https://www.grants.gov/. It contains information about US government-funded programs and projects.} (called {\em Grants} henceforth), a website containing a list of research proposals from several different agencies. We found that the XML archives had nothing linking them back to the original grant page, and did not include active grants, so we did not use them.  The Grants search page had a wider variety of organizations to choose from, and included active grants updated daily. However, the information was both inconsistent and costly to parse, as getting from the search to the Grants page took inordinate amount of processing time to exhaustively search the list.

To get potential team members, a list of researchers is made by pulling information from a University College's  faculty list\footnote{https://sc.edu/study/colleges\_schools/engineering\_and\_computing/
faculty-staff/index.php} hosted on their web site. This list is then filtered down to a subset in research role using heuristics that is applied to
 job titles and designations. To accurately filter and extract our data for these researchers, we perform two additional check. First, we check the individual's university webpage for keywords, then we search Google Scholar API\footnote{Using library scholarly - https://pypi.org/project/scholarly/.} for the individual's scholar page. From these sources, we are able to retrieve past published works, and a list of keywords of the individual's technical skills (research interests). We then consolidate and normalize skills obtained from different sources using NLP techniques - 
stop word removal, stemming and lemmatization.
 When no skill information is available for a person, they are omitted from the system. 
In numbers, for the current prototype, 318 total experts were extracted from the University's employees' websites out for which 
78  were removed on the basis of designation. In the remaining 240,
205 had either research background from University website or interests as retrieved from Google Scholar. 83 actually had Google Scholar profiles and were confirmed researchers at the University eligible to apply for calls.
In Table~\ref{tab:extr-results}, Users/Research refers to the extraction of researcher information.

We also retrieve previous awards by funding agencies to analyze features and trends in successful proposals. We focused on NSF and NIH who are front-runners for  research funding in the US with NSF having budgeted
\$6.5B\footnote{https://www.nsf.gov/pubs/2020/nsf20002/pdf/nsf20002.pdf} in FY
2019 and the NIH
\$31.8B\footnote{https://report.nih.gov/nihdatabook/report/283}. Previous award data was obtained by accessing the agencies’
websites\footnote{https://www.nsf.gov/awardsearch/download.jsp} and downloading ZIP files containing XML files
corresponding to all projects (and subprojects) awarded across  years of interest. We do not use previous award data in current prototype but could use in future.

\subsection{Matching Proposal Calls  To Researchers Skills}

We now describe how a proposal call (RFP) is matched to researchers' skills representing teaming demand with potential supply. 
We experimented with multiple matching methods all of which use  the call synopsis obtained by CE and researchers' skills. 

As the first method, we consider approximate string matching as implemented in the Fuzz library\footnote{https://github.com/seatgeek/thefuzz}. We will refer to it as fuzzy. The second method we use is embedding based.
For this, we rely on SPECTER (Scientific Paper Embedding using Citation Informed Transformers\footnote{https://github.com/allenai/specter})~\cite{specter2020cohan}. SPECTER is a representation learning method based on BERT \cite{bert} but uses a powerful signal of document-level relatedness: the citation graph. SPECTER is particularly helpful for recommendation tasks with reported performance like nDCG (Normalized Discount Cumulative Gain\footnote{https://www.geeksforgeeks.org/normalized-discounted-cumulative-gain-multilabel-ranking-metrics-ml/} index of 53.9). 
We first generate  document-level embedding on the collected corpus of call synopsis  using SPECTER. Then, we implement a match function that takes call synopsis and a skills vector and provides score based on the computed embedding.

As an example, on call summary - 
{\em The Artificial Intelligence and Cognitive Science AICS program focuses on advancing the state of the art in Artificial Intelligence and Cognitive Science. The program supports research and related education activities fundamental to the development of computer systems capable of performing a broad variety of intelligent tasks and to the development of computational models of intelligent behavior across the spectrum of human intelligence}, and a  skill profile = ['Artificial Intelligence', 'Services', 'Smarter Cities (Water', 'Health', 'Traffic)'],  the fuzzy method's matching score is 51 on a 0-100 scale where 100 represents full match. For the same, the  embedding matching score is 79 on a 0-100 scale where 100 represents full match. 
The call is highly relevant to the researcher and the embedding method reflects this more accurately than the fuzzy method. We note that the two scores are from two different systems, and in theory, may be incomparable. So, to verify whether they correspond to what researchers perceive in practice, we created a large subset of matches with both methods and gave to researchers for their evaluation. As reported in evaluation section, we found empirically that researchers preferred matches which scored higher by embedding based method. Hence, we use it in our implementation.

For the researcher shown in Figure~\ref{fig:user-login}, the system found 10 calls that have the highest match. These are used to explore possible teams with the researcher as team leader.

\subsection{Team Suggestions}

To get team recommendations from matches, we use a greedy strategy for team formation. First, we run our matching algorithm in a user-item recommendation format i.e. for each user system recommends at most $k$ (= 10) relevant synopsis (of a RFP) in descending order of match score. 
Then, in second step, we form {\em team groups} to associate users with recommended RFP based on short and long term constraints.
The constraints we use have been obtained after discussions in focus groups with University administrators and researchers. They are: (a) limit upper team size to 5 members unless budget size allows larger participants, (b) have at least  \$50K per participant (c) each member should have at least one non-overlapping skill with another. The approach can easily incorporate new constraints and can explain request for change to team composition by showing which constraints would be respected or violated. We anticipate additional constraints at user-level (by researchers) like in maximum number of recommendations in a time period.
If the number of users recommended for a call are more than team size limit, we sort the matched users and select in the descending order of matched score. 
At the end of this step, we get user-specific recommendation of calls and possible teams, and call-specific recommendation of potential teams. 

We present an example of team formed by the above method in Table~\ref{tab:team-results} consisting of 5 researchers on a historical call. 
In ULTRA screen, the user could go to the link associated with this RFP, the synopsis obtained is from the section of call - \emph{Summary of Program Requirements Section}. It is related to computational neuroscience and computer science research. The team has 4 researchers directly related to {\em bio-informatics} or {\em neuromorphic computing} and the fifth one related to {\em deep learning}. Therefore, we see that system is able to match very effectively and its results are promising.
\begin{table}
\centering
    \begin{tabular}{|p{2cm}|p{4cm}|p{1cm}|  }
\hline
    
          User & Research Interests & Score   \\ \hline 
     {Agostinelli Forest}  & ['Artificial Intelligence, Deep Learning, Reinforcement Learning, Search, Bioinformatics'] &84
  \\ \hline
     Hu Jianjun & ['Deep learning, machine learning, materials informatics, bioinformatics, evolutionary computation'] & 84
 \\ \hline 
     
     Valafar Homayoun & ['Computational Biology, Bioinformatics, Artificial Intelligence, Machine Learning, Neural Information Processing']& 83
    \\ \hline
     Zand Ramtin & [‘Hardware Design for Machine Learning Systems , Neuromorphic Computing , Emerging Nanoscale Electronics’]& 80
  \\ \hline  
     Luo Lannan & ['Sofware and Systems Security, Mobile Security, Software Engineering, Programming Language, Deep Learning']& 76
  \\ \hline
    \end{tabular}
    \caption{Team Formation Result generated by ULTRA for historic NSF call - Innovative Approaches to Science and Engineering Research on Brain Function. \small{URL: https://www.nsf.gov/pubs/2014/nsf14504/nsf14504.htm}}
    \label{tab:team-results}
\end{table}
\section{Evaluation}

We have conducted preliminary evaluation ULTRA's output.
The main output of the system is a list of experts who compose a team and whose skill set, experience and scientific excellence are compatible with the RFP topic.
The evaluation of the effectiveness of the system was done retrospectively by analyzing historic data of both the previous RFPs and experts’ engagement with the latter.

During initial evaluation,  seven users from a College in the University 
were selected for evaluation. The aim of the evaluation was to determine if the system had identified the RFPs for each user fitting well within their competence set. For each user the system had generated ten best matches, based on the pool of historic RFPs. The RFP match was assessed in Likert scale 1...10, by the user themselves or by the expert panel, if the user feedback was not reached. The human assessment revealed the following: for two users, all the ten system matched RFPs fit very well within their competence (at least rating 7 in Likert scale). For the remaining five users, nine out of ten RFPs were matched very well, whereas one RFP had been assessed as 6 or lower on Likert scale. Altogether, from 70 matches, 65 RFPs were assessed by respondents as 7 or higher in Likert scale, demonstrating the high effectiveness of the system in matching users' skills to historic RFPs.

In particular, the user\footnote{Jorge Crichigno, PhD, Associate Professor in Integrated Information Technology} in Figure~\ref{fig:user-login}
 commented that a new search term {\em cybertraining} helped him to locate a RFP which would have normally be  unknown to him. Also, he noted that ``I looked at the NSF grants and they seem very relevant", noting the satisfaction with ULTRA's output.

Next, we obtained the records of awarded RFPs made available by the University's division responsible for sponsored research.  We manually annotated the synopsis for each awarded project. The included awards were limited to NSF data, as information about the awards is publicly well accessible. We ran our user-item recommendation system for the corpus of synopsis. We got the recommendation for Principal Investigators in the form of 10 best awards - the results were compared to actual award data, resulting in the accuracy of 54.5 per cent where the system was able to match the users with the actual awards they had received. This level can be considered satisfactory system output, as the success in getting the awards does depend also on many other factors, besides competence matching. In future, we will incorporate matching and teaming enhancements  to improve over the matching baselines.

\section{Ultra Family of Tools}

In this section,  we will describe some of the important tools that we have developed as part of the Ultra effort. They started out as useful features that we then made into  stand-alone capabilities recognizing their potentia for wider usage.

\subsection{Tools for Exploring Ultra Data}

We have created a webpage where one can see or retrieve all the content we have extracted about the users, proposal calls, and previous awards by selecting through the dropdown. When one chooses the {\em Proposal calls} or {em Users} in the dropdown, one can search through the proposals using the {\em proposal id}, {\em agency id}, and users using {\em user id} and {\em username}. 
Additionally, we have   Application Programming Interface (API) documentation  of how one can retrieve records of proposals, users, or awards. And also, we can extract the desired records by giving the {\em agency id} for proposals, {\em username} for users, and {\em award number} for the awards, respectively. 

\subsection{Unsupervised Text Visualization with KITE}

\begin{figure*}[t]
\centering
\caption{Using \emph{KITE} to explore the webpage (document) of a Researcher}.
\label{fig:kite}
\fbox{\includegraphics[width=\linewidth, height = 8cm]{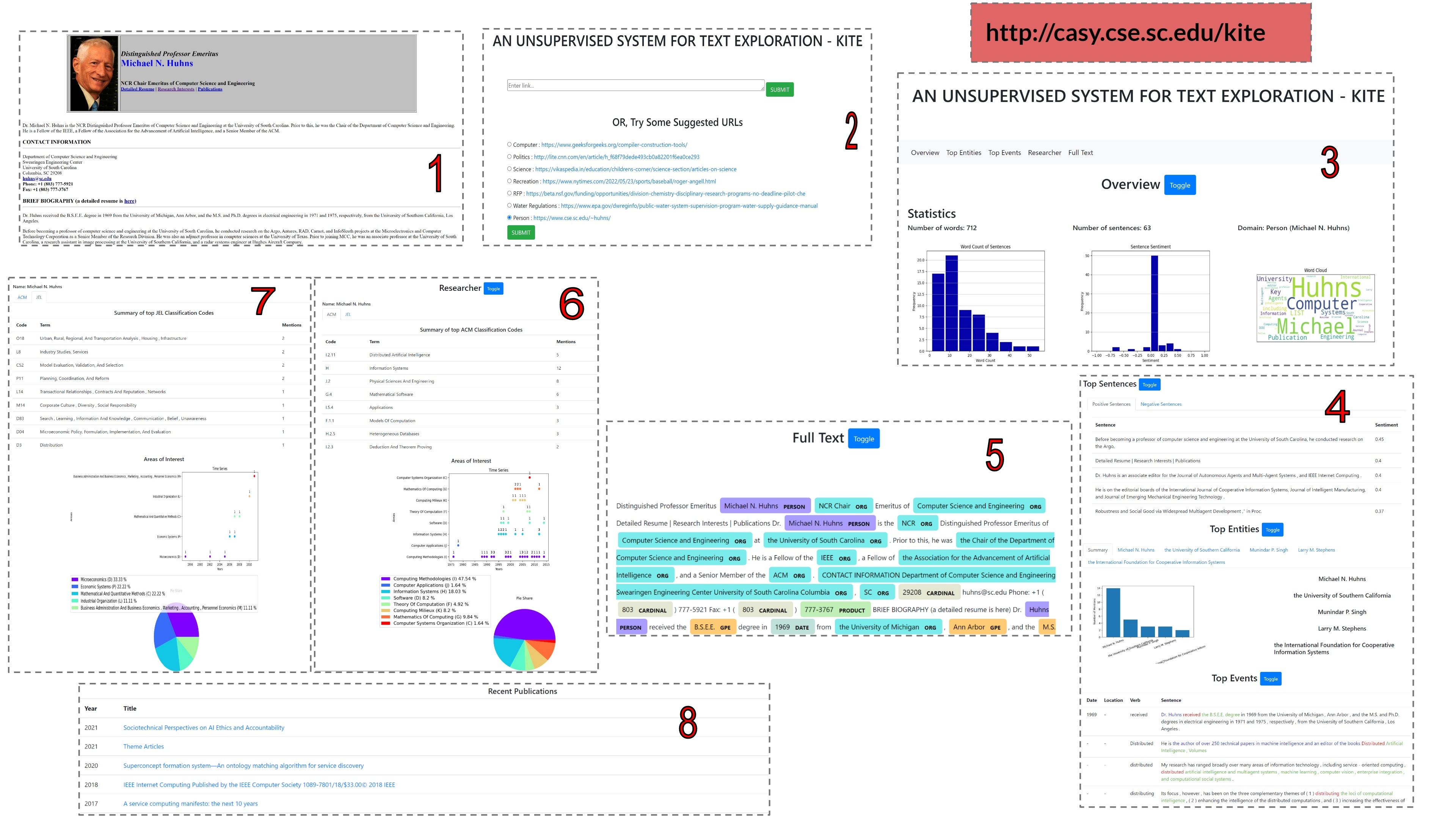}}
\end{figure*}

Kite \cite{kite} is an unsupervised system for exploring textual data and generating insights. This is used for unstructured documents like news articles, survey results, profiles of people and proposals. In Ultra case, one can use KITE to explore the background of researchers and proposal calls that we extract. Figure~\ref{fig:kite} shows an example of it use for a researcher. 
Here, View-1 displays the sample URL of the person's webpage, View-2 displays the user interface of KITE, and Views3-8 shed light about the person's technical background using concepts, relationships and their change over time, including  using standard terminologies  from the computing (ACM) and  economics
(JEL) domains.


\subsection{Text to Classsification Mapper}
\begin{figure}
\centering
\caption{Text to Classification Mapper}.
\label{fig:mapper}
\includegraphics[width=\linewidth]{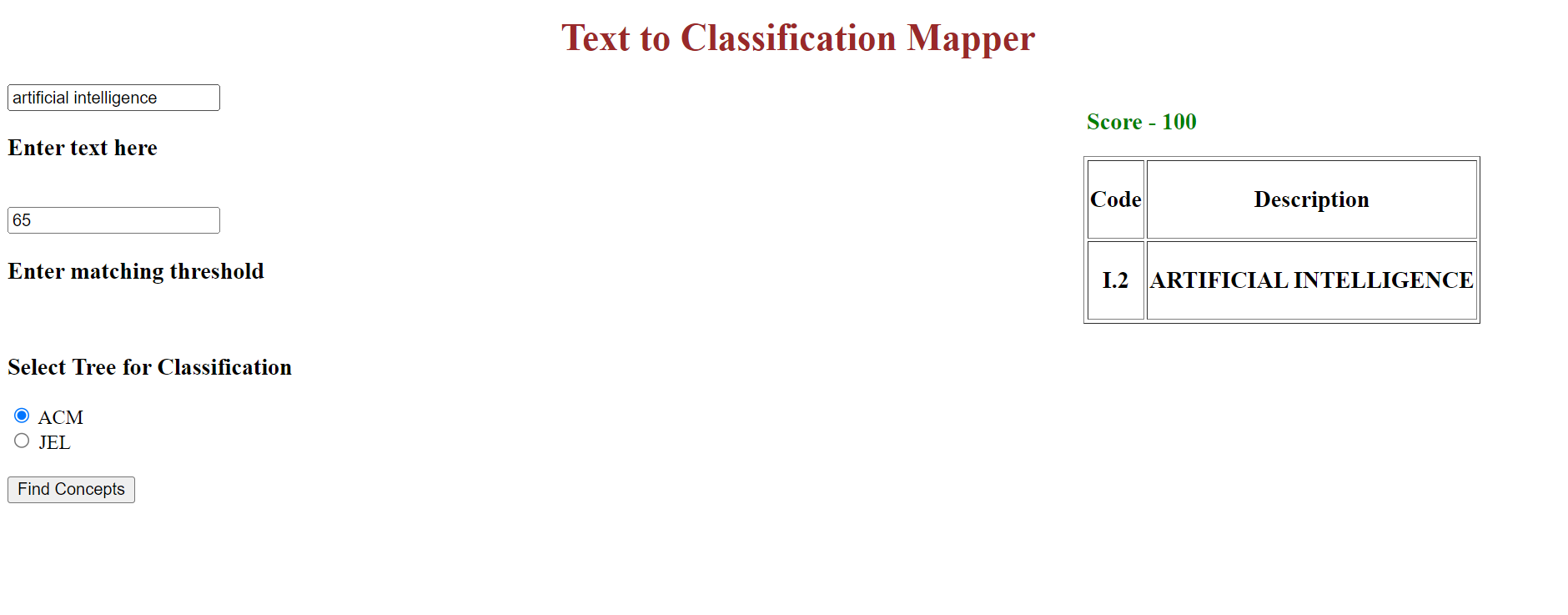}
\end{figure}

Mapper is a tool that takes the input as a text and matching threshold as a number and returns the ACM or JEL classification codes and description based on the input text. Figure~\ref{fig:mapper} shows an example of how to get the result in a table using the input text.

\section{Discussion and Future Work}

We now discuss the impact of our approach, initial user feedback and opportunities for future improvement.
\noindent {\bf Initial Feedback:}
The feedback we have received from administrative and research users has been encouraging.
One is to expand ULTRA's data coverage beyond engineering. Here, {\em Web of Science} has wider coverage about researchers than Google Scholar. Forming teams is challenging, especially for cross disciplinary teams like social sciences, arts, humanities.
Second, users want the system to assign 
star rating to team leaders who have had good experiences and track record. Third, some proposals expect junior faculty or member from disadvantaged groups to be in the lead, and this could be prioritized by the tool for teaming. Fourth,  sometimes a user displays having competence in certain fields, but this is not reflected on their track record, such as previous publications and projects. The system can then use reliable online information to properly characterize users' competence. At the same time, the system should also allow people to update their research profiles. 
Overall, stakeholders agree that this is a complex problem and our system is trying to get ahead of curve trying to solve this problem.

\noindent {\bf Future Directions:}
The current prototype can be extended along many development directions. 
First, we can increase the scale of experiments and data about researchers to outside of engineering.
Second, we can normalize the technical areas that are used from the proposal call and researcher profiles beyond NLP techniques.
In \cite{kite}, the authors present an unsupervised approach to visualize text documents where they normalize skills of researchers and those mentioned in technical calls. In our tool, we can use their approach and focus on the area of Computing to use the ACM Computing Classification System (1998) \cite{acm-class,class-article} - a set of 1532 \emph{Terms} and their \emph{Codes}. Beyond ACM, one can use American Economic Association  Classification for the field of economics and the healthcare classification ICD-10 containing codes for diseases, signs and symptoms, abnormal findings, complaints, social circumstances, and external causes of injury or diseases.
Third, one can improve matching using Bandit and Hedonic games methods. The data about previous awards could be useful in this regard. Fourth, one can also improve team formation approach with metrics for short and long-term optimization of team proposal results and performance on funded projects. 

Finally, it is an open question whether  users will actually follow an Artificial Intelligence (AI) system's recommendations for collaboration even if there is analytically a higher chance of success. After all, the AI system can only be as accurate as the data it uses and there are many intricacies of interpersonal collaboration that data may fail to capture. Only a careful long-term  longitudinal evaluation will be able to establish if data-driven teaming recommendations actually works and we are only in the early stages of this investigation. If that is the case,  this area would be a fertile ground to nudge {\em groups of} users and personalize their  recommendations towards high potential collaboration.


\section{Conclusion}

In this paper, we tackled the emerging problem of how AI can assist researchers identify right funding opportunities, assessing their suitability for the program and suggest teaming opportunities for successful proposals. This is an example of the general problem of team formation when opportunities may repeat over time. We proposed an approach 
based on content extraction from 
open sites about calls and researchers, matching users to calls, and then forming teams based on business constraints. We described a prototype, ULTRA, 
and presented preliminary empirical evidence that the approach is promising. This work lays the basis for future work on AI-assisted teaming spanning multiple disciplines at University-scale. 


.

\section{\label{sec:ack}
 Acknowledgement}

We will like to thank Ronak Shah and Austin Hetherington for their contributions in early implementation.


\bibliographystyle{IEEEtran}
\bibliography{bibs/references}


\end{document}